# Alternating direction method of multipliers applied to medical image restoration

Kenya Murase[1,2]

[1]Department of Medical Physics and Engineering, Faculty of Health Science, Graduate School of Medicine, Osaka University, Suita, Osaka, Japan

[2]Department of Future Diagnostic Radiology, Graduate School of Medicine, Osaka University, Suita, Osaka, Japan

**Abstract**

We investigate the effects of the regularization parameter for the $\ell_1$ norm ($\alpha_1$) and penalty parameter ($\mu$) in the alternating direction method of multipliers (ADMM) on the quality of restored medical images. Simulation studies are performed using images degraded by a point spread function (PSF) and Gaussian noise. The *j*-th column of the system matrix ($A$) is calculated by convolving the image with unity at pixel *j* and zero at all other pixels and the PSF. The simulation studies show that the mean structural similarity index is maximal when $\mu/\mu_0$ is approximately 10 to 20, where $\mu_0 = 1/\|A^T g\|_\infty$, with $A^T$ and $g$ being the transpose of $A$ and the observed data, respectively. The restored image became blurred with a decrease in $\alpha_1$. This study will be useful for identifying optimal parameter values in the ADMM when applied to medical image restoration.

**Keywords:** Alternating direction method of multipliers, Medical image restoration, System matrix, Point spread function, Structural similarity index

# 1 Introduction

Because medical images are typically degraded by the finite spatial resolution of imaging modalities and corrupted by noise, image restoration is required to improve these images and enhance their diagnostic accuracy [1]. The aim of image restoration is to obtain the best estimate of an object using the information in a blurred image contaminated by noise.

Image restoration is one of the most classical linear inverse problems in image processing. With the emergence of compressed sensing (CS), these studies have regained attention [2]. In signal and image processing and CS problems, sparsity-inducing regularizers such as the $\ell_1$ norm and total variation (TV) norm have often been used effectively for enforcing sparseness and piece-wise smoothness, respectively [3, 4].

The alternating direction method of multipliers (ADMM) was developed in the 1970s [5] and belongs to the family of augmented Lagrangian techniques [6]. The ADMM has been applied in many areas, including signal and image processing [7–9], statistics and machine learning [10], and system control [11]. In 2011, Afonso et al. [6] developed a fast method for solving constrained optimization problems using variable splitting [12] and the ADMM, which is known as the constrained split augmented Lagrangian shrinkage algorithm (C-SALSA). They reported that the C-SALSA is faster than state-of-the-art methods, such as the fast iterative shrinkage/thresholding algorithm and Nesterov's algorithm [6].

When solving optimization problems, the optimal regularization parameter values are typically searched by trial and error using hand tuning. When the number of regularizers and hence the number of parameters to be optimized are significant and/or the problem is ill posed, considerable time and effort is required to achieve their optimal values, particularly when they are searched randomly. If a parameter that can be used as a merkmal and reduces the search range exists, then the search time and effort will be reduced.

When applying the ADMM to image restoration, a linear imaging system is assumed, and a matrix operator related to an underlying source image and observed data is used [6, 12]. Herein, we refer to the matrix operator as the system matrix. Recently, we developed an image-restoration method using a simultaneous algebraic reconstruction technique based on a point spread function (PSF) [1]. In this method, the relationship between the elements of the system matrix and the PSF is derived and used [1].

The purpose of this study is to investigate the usefulness of the ADMM when applied to medical image restoration through simulation studies. Furthermore, we present a method for generating the system matrix in the ADMM using the PSF and introduce a parameter that can be used as a merkmal for identifying the optimal parameter values. Additionally, we investigated the effects of regularization and penalty parameters in the ADMM on the image quality of restored images.

## 2 Materials and methods

### 2.1 Alternating direction method of multipliers

A linear imaging system can be modeled as follows:

$$\boldsymbol{Af} = \boldsymbol{g}, \tag{1}$$

where $\boldsymbol{A}, \boldsymbol{f}$, and $\boldsymbol{g}$ denote the system matrix, underlying source image vector, and observed image vector, respectively. The aim of image restoration is to estimate $\boldsymbol{f}$ from $\boldsymbol{g}$. This can be reduced to solving the following optimization problem:

$$\min_{\boldsymbol{f}}\{\alpha_1\|\boldsymbol{f}\|_1 + \alpha_2 TV(\boldsymbol{f})\} \text{ subject to } \|\boldsymbol{Af} - \boldsymbol{g}\|_2 \leq \varepsilon, \tag{2}$$

where $\|\boldsymbol{f}\|_1$ and $TV(\boldsymbol{f})$ denote the $\ell_1$ norm and TV norm of $\boldsymbol{f}$, respectively. In this study, the following isotropic TV norm is used as $TV(\boldsymbol{f})$:

$$TV(\boldsymbol{f}) = \sum_x \sum_y \sqrt{(f_{x+1,y} - f_{x,y})^2 + (f_{x,y+1} - f_{x,y})^2}, \tag{3}$$

where $f_{x,y}$ denotes the image intensity at pixel (x, y) [13]. The symbols $\alpha_1$ and $\alpha_2$ in Eq. (2) are the regularization parameters for the $\ell_1$ and TV norms, respectively. In this study, we assumed $\alpha_1 + \alpha_2 = 1$. The symbol $\varepsilon \, (\geq 0)$ is a parameter that depends on the noise variance.

When using variable splitting [12] and the ADMM, $\boldsymbol{f}$ is computed iteratively as follows [6]: Beginning with the initial estimates $\boldsymbol{f}^1$, $\boldsymbol{v}_k^1$, and $\boldsymbol{d}_k^1$ ($k = 0,1,2$), an image vector ($\boldsymbol{f}^n$) and multipliers [$\boldsymbol{v}_k^n$ and $\boldsymbol{d}_k^n$ ($k = 0,1,2$)] at each iteration $n$ are updated as [6]

$$\boldsymbol{f}^{n+1} = \left[\sum_{k=0}^{2} \boldsymbol{H}_k^T \boldsymbol{H}_k\right]^{-1} \left[\sum_{k=0}^{2} \boldsymbol{H}_k^T (\boldsymbol{v}_k^n + \boldsymbol{d}_k^n)\right], \tag{4}$$

$$\begin{aligned}\boldsymbol{v}_k^{n+1} &= \arg\min_{\boldsymbol{v}} \left\{\Phi_k(\boldsymbol{v}) + \frac{\mu}{2}\|\boldsymbol{v} - \boldsymbol{H}_k\boldsymbol{f}^{n+1} + \boldsymbol{d}_k^n\|_2^2\right\} \\ &= \Psi_{\Phi_k/\mu}(\boldsymbol{H}_k\boldsymbol{f}^{n+1} - \boldsymbol{d}_k^n) \; (k = 0,1,2),\end{aligned} \tag{5}$$

and

$$\boldsymbol{d}_k^{n+1} = \boldsymbol{d}_k^n - \boldsymbol{H}_k\boldsymbol{f}^{n+1} + \boldsymbol{v}_k^{n+1} \; (k = 0,1,2). \tag{6}$$

Here, $T$ denotes the transpose of a matrix; $\boldsymbol{H}_0 = \boldsymbol{A}$ and $\boldsymbol{H}_1 = \boldsymbol{H}_2 = \boldsymbol{I}$, where $\boldsymbol{I}$ is an identity matrix. In this study, $\boldsymbol{f}^1$ was set to a uniform image vector with zero pixel intensity, and $\boldsymbol{v}_k^1$ and $\boldsymbol{d}_k^1$ ($k = 0,1,2$) were set to zero. $\Psi_{\Phi_k/\mu}(\boldsymbol{x})$ in Eq. (5) is the Moreau proximal mapping of $\Phi_k(\boldsymbol{x})$ [6].

For $k = 0$, $\Phi_0(\boldsymbol{x})$ in Eq. (5) is expressed by the 0-infinity indicator function $[\iota_{E(\varepsilon,\boldsymbol{I},\boldsymbol{g})}(\boldsymbol{x})]$ of the set $\{\boldsymbol{x}|\ \|\boldsymbol{x} - \boldsymbol{g}\|_2 \leq \varepsilon\}$, and its Moreau proximal mapping is expressed as [6]

$$\Psi_{\iota_{E(\varepsilon,\boldsymbol{I},\boldsymbol{g})}/\mu}(\boldsymbol{x}) = \boldsymbol{g} + \begin{cases} \varepsilon \dfrac{\boldsymbol{x}-\boldsymbol{g}}{\|\boldsymbol{x}-\boldsymbol{g}\|_2} & \text{for } \|\boldsymbol{x} - \boldsymbol{g}\|_2 > \varepsilon \\ \boldsymbol{x} - \boldsymbol{g} & \text{for } \|\boldsymbol{x} - \boldsymbol{g}\|_2 \leq \varepsilon \end{cases}. \tag{7}$$

For $k = 1$, $\Phi_1(\boldsymbol{x})$ is expressed by the $\ell_1$ norm in Eq. (2), i.e., $\Phi_1(\boldsymbol{x}) = \alpha_1 \|\boldsymbol{x}\|_1$, and its Moreau proximal mapping is expressed by the soft thresholding function as $\Psi_{\Phi_1/\mu}(\boldsymbol{x}) = S_{\Phi_1/\mu}(\boldsymbol{x}, \alpha_1/\mu) = \text{sign}(\boldsymbol{x}) \cdot \max(\lfloor \boldsymbol{x} \rfloor - \alpha_1/\mu, 0)$ [6]. For $k = 2$, $\Phi_2(\boldsymbol{x})$ is expressed by the TV term in Eq. (2), i.e., $\Phi_2(\boldsymbol{x}) = \alpha_2 TV(\boldsymbol{x}) = (1 - \alpha_1)TV(\boldsymbol{x})$. In this study, the Moreau proximal mapping of $\Phi_2(\boldsymbol{x})$ $[\Psi_{\Phi_2/\mu}(\boldsymbol{x})]$ was solved using the fast gradient projection method developed by Beck et al. [4]. The symbol $\mu$ ($\geq 0$) in Eq. (5) denotes the penalty parameter.

The iterative procedure above was repeated until $\|\boldsymbol{f}^{n+1} - \boldsymbol{f}^n\|_2/\|\boldsymbol{f}^n\|_2 < \varepsilon_{tol}$ was satisfied or the maximum number of iterations ($N_{max}$) was reached. In this study, $\varepsilon_{tol}$ and $N_{max}$ were set to $10^{-4}$ and 100, respectively. When $\boldsymbol{f}^{n+1}$ in Eq. (4) contained negative values, these values were set to zero.

## 2.2 Calculation of system matrix elements

When considering the application of the ADMM to planar imaging, element $\boldsymbol{A}$ in Eq. (1) ($a_{ij}$) represents the contribution of the source pixel $j$ to the observation pixel $i$ and is expressed as

$$a_{ij} = S(i) \cdot PSF(i - j), \tag{8}$$

where $S(i)$ denotes the sensitivity of an imaging device at pixel $i$, and $PSF(i - j)$ is the PSF, which is expressed in terms of the spatial offset between observation pixel $i$ and source pixel $j$ [1]. In this study, the PSF was normalized to the unit sum, $\sum_j PSF(i - j) = 1$.

When using Dirac's delta function, $a_{ij}$ can be expressed in integral form as follows:

$$a_{ij} = S(i) \int_{-\infty}^{\infty} \delta(j' - j) \cdot PSF(i - j')dj'. \tag{9}$$

As shown in Eq. (9), $a_{ij}$ can be expressed as the intensity at pixel $i$ of the image obtained by convolving $\hat{f}_j$ and the PSF, multiplied by $S(i)$, where $\hat{f}_j$ represents the image vector with unity at pixel $j$ and zero for all other pixels.

**2.3 Simulation study using numerical phantom**

To perform the simulation studies, a numerical phantom was constructed as follows: First, we generated a 128 × 128 pixel image, including gray matter (GM), white matter (WM), and cerebrospinal fluid (CSF), by segmenting the brain images of healthy participants obtained via magnetic resonance imaging [14, 15]. We assumed that the cerebral blood flow values in the GM, WM, and CSF were 60 mL/100 g min$^{-1}$, 20 mL/100 g min$^{-1}$, and 0 mL/100 g min$^{-1}$, respectively [1]. Next, the effect of the finite spatial resolution was considered by convolving the numerical phantom image generated above with the PSF. In this study, we assumed that the PSF had a circularly symmetric Gaussian distribution and its full width at half maximum (FWHM) was 6 mm (4 pixels) [1]. In addition, Gaussian noise was added to the image generated above using normally distributed random numbers with zero mean and unit variance, as described in our previous paper [1]. The noise levels were assumed to be 2%, 5%, and 10%, where the noise level was defined as the percentage of the noise standard deviation divided by the maximum pixel intensity of the degraded image. Figures 1(a) and 1(b) show the numerical phantom image and degraded image with a noise level of 10%, respectively.

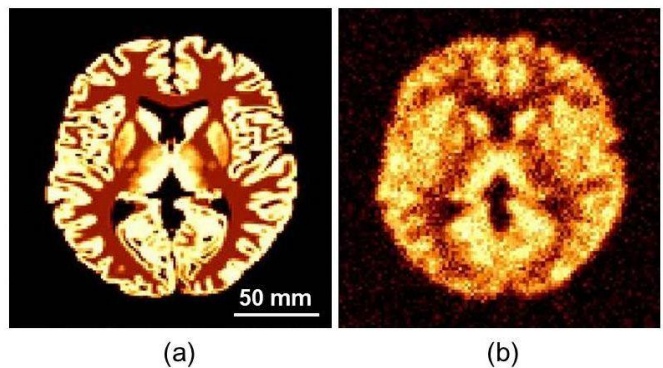

(a)                            (b)

**Fig. 1** (a) Numerical phantom image. Scale bar = 50 mm. (b) Degraded image with noise level of 10%.

In the simulation studies, $S(i)$ in Eq. (8) was assumed to be unity, and the elements of $A$ were calculated using Eq. (9). The symbol $\varepsilon$ in Eq. (7) was calculated from $\varepsilon = \sqrt{N_{ob}} \cdot \sigma_n$, where $N_{ob}$ and $\sigma_n$ denote the number of observations and noise standard deviation, respectively. In this study, $N_{ob}$ represents the total number of pixels, i.e., $128 \times 128$, and $\sigma_n$ was 1.2, 3.1, and 6.8, for 2%, 5%, and 10% noise levels, respectively. The number of iterations required to solve Eq. (5) for TV minimization using the fast gradient projection method [4] was set to 20 at each iteration *n*.

## 2.4 Evaluation

The restored images obtained by the ADMM were quantitatively evaluated using two measures. One was the percent root mean square error (PRMSE), and the other was the mean structural similarity index (mSSIM) proposed by Wang et al. [16]. The details of these measures are available in our previous paper [1].

## 3 Results

The PRMSE values for $\alpha_1$ values of 0.2, 0.5, and 0.8, are shown as a function of $\mu/\mu_0$ in the upper row of Fig. 2, whereas the mSSIM values are shown in the lower row of Fig. 2. The left, middle, and right columns of Fig. 2 show cases with noise levels of 2%, 5%, and 10%, respectively. As shown in the upper row of Fig. 2, the PRMSE values for noise levels of 5% and 10% were minimal at $\mu/\mu_0 \approx 20$. The $\mu/\mu_0$ value at which the PRMSE was minimal tended to decrease with increasing noise level. As shown in the lower row of Fig. 2, when $\alpha_1$ was 0.5 or 0.8, the mSSIM peaked at $\mu/\mu_0$ of approximately 10 to 20, and the peaks became clearer with increasing noise level. When $\alpha_1$ was 0.2, the mSSIM increased with $\mu/\mu_0$ to approximately 20 and decreased slowly or plateaued thereafter. It is noteworthy that the PRMSE and mSSIM for the degraded images with noise levels of 2%, 5%, and 10% were 26.9% and 0.762, 28.1% and 0.703, and 32.3% and 0.577, respectively.

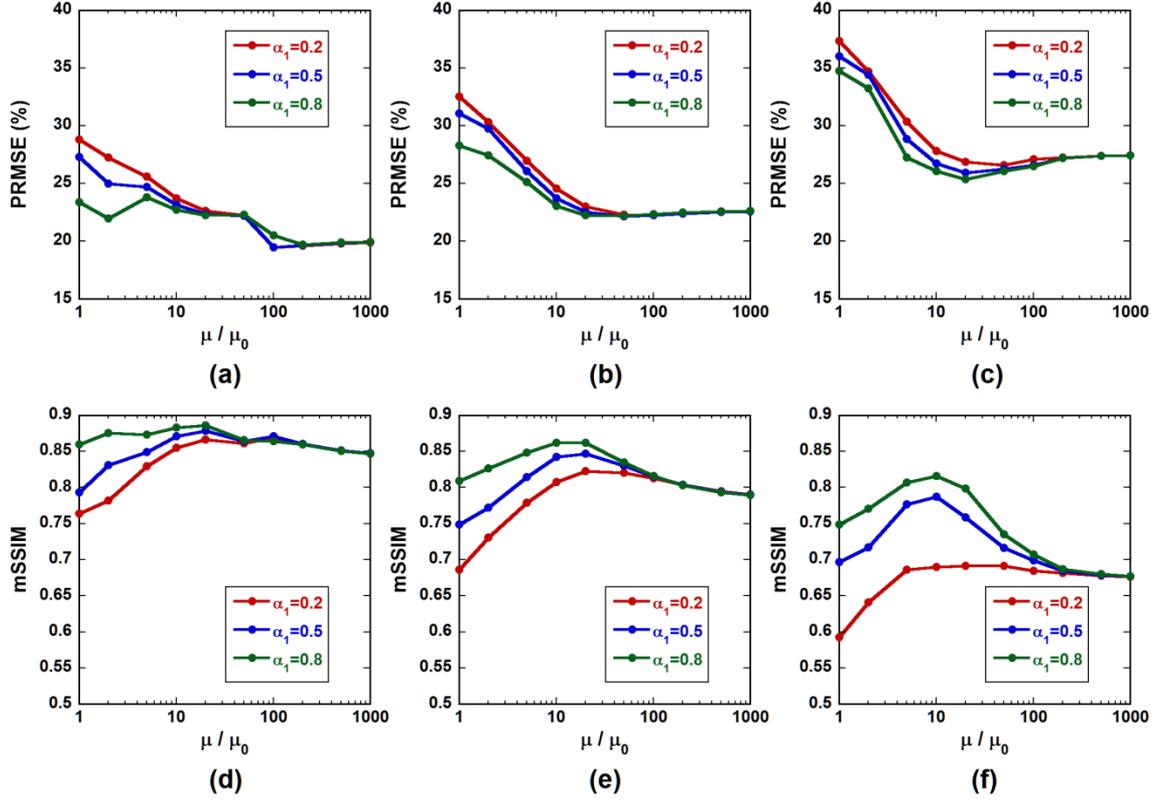

**Fig. 2** (Upper row) Percent root mean square error (PRMSE) as a function of $\mu/\mu_0$ for various $\alpha_1$ values. (a), (b), and (c) show cases with noise levels of 2%, 5%, and 10%, respectively. (Lower row) Mean structural similarity index (mSSIM) as a function of $\mu/\mu_0$ for various $\alpha_1$ values. (d), (e), and (f) show cases with noise levels of 2%, 5%, and 10%, respectively.

Figure 3 shows the restored images of the numerical phantom [Fig. 1(a)] obtained using the ADMM with $\mu/\mu_0$ values of 1, 10, 20, 50, and 100 (from left to right columns). The upper, middle, and lower rows in Fig. 3 show cases with $\alpha_1$ values of 0.2, 0.5, and 0.8, respectively. In these cases, the noise level was assumed to be 10%. A degraded image is shown in Fig. 1(b). As shown in Fig. 3, the spatial resolution improved with increasing $\alpha_1$, particularly when the $\mu/\mu_0$ value was 1 or 10. When $\alpha_1$ was constant, the spatial resolution improved as $\mu/\mu_0$ increased to approximately 20, and it did not change significantly thereafter. When $\alpha_1$ was 0.2 or 0.5, this observation was more evident compared with when $\alpha_1$ was 0.8.

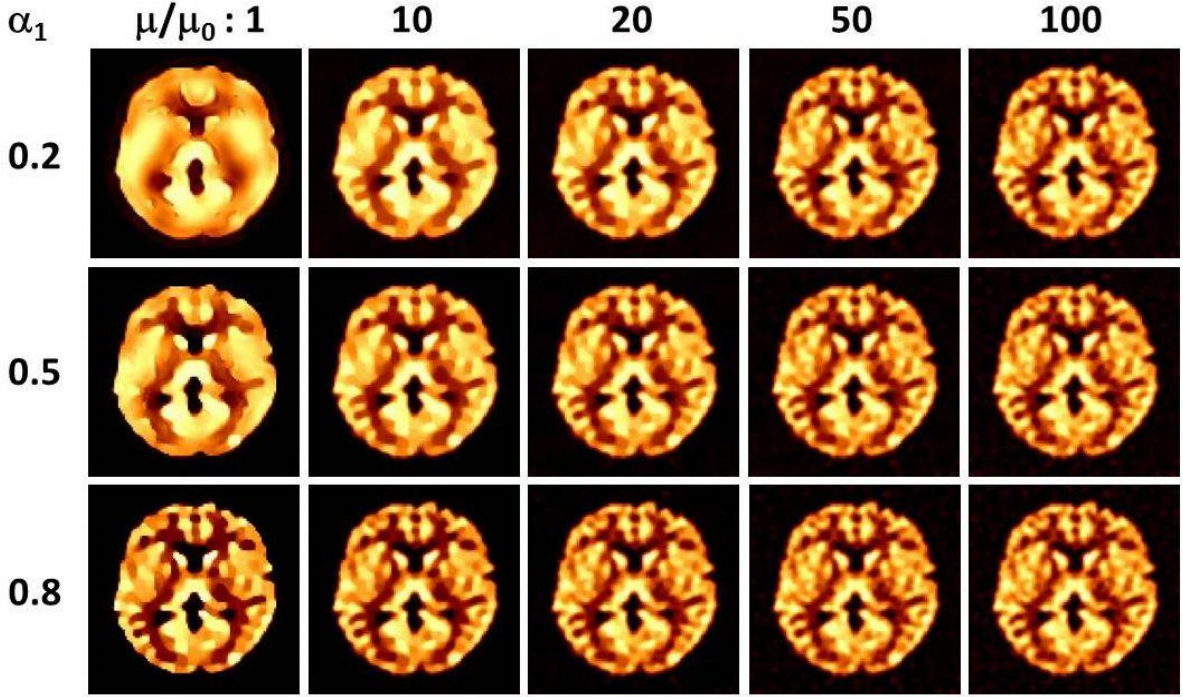

**Fig. 3** Restored images of numerical phantom [Fig. 1(a)] obtained using alternating direction method of multipliers (ADMM) with μ/μ$_0$ values of 1, 10, 20, 50, and 100 (from left to right columns). Upper, middle, and lower rows show cases for α$_1$ values of 0.2, 0.5, and 0.8, respectively. In these cases, the noise level was assumed to be 10%.

## 4 Discussion

In this study, we applied the ADMM to medical image restoration, and investigated its usefulness through simulation studies. Our results demonstrate the usefulness of the ADMM. In the ADMM, it is assumed that the relationship between the source image and observed data is as expressed in Eq. (1), and the system matrix that connects them, i.e., $A$, is known. Herein, we present a method for calculating the elements of $A$ using the PSF [Eq. (9)]: If the PSF and the sensitivity of an imaging device do not change, $A$ must be calculated only once, and the obtained $A$ can be saved and recalled when required.

When applying the ADMM, the optimal values of the regularization parameters must be identified. The tuning parameters used in this study were $\alpha_1$ and $\varepsilon$ [as shown in Eq. (2)] and $\mu$ in Eq. (5). It is noteworthy that $\alpha_2$ in Eq. (2) can be determined from $1 - \alpha_1$. Identifying these optimal values is not straightforward and is typically performed by trial and error using hand tuning, which involves considerable time and effort. In this study, we introduced $\mu_0$ ($=1/\|A^T g\|_\infty$) as a merkmal and investigated the optimal parameter values based on $\mu_0$. The term $\|A^T g\|_\infty$ is the maximum of the critical solution of $\min_f \left\{\frac{1}{2}\|Af - g\|_2^2 + \lambda\|f\|_1\right\}$ at $f = 0$, where $\lambda$ denotes the regularization parameter for the $\ell_1$ norm [10]. Our results (Figs. 2 and 3) suggest that this parameter is beneficial for identifying the optimal $\mu$ value from many candidates using trial and error, which will reduce the time and effort required for optimization.

When investigating the optimal parameter values in the simulation studies, we used two quantitative measures, i.e., the PRMSE and mSSIM (Fig. 2). As shown in Fig. 2, although some exceptions were observed, particularly when $\alpha_1$ was 0.2 and the noise level was 2%, minimum and maximum values of PRMSE and mSSIM were obtained, respectively, when $\mu/\mu_0$ was 10–20. Boyd et al. [10] reported that the number of iterations required to solve a single least absolute shrinkage and selection operator problem was minimal when the regularization parameter described above ($\lambda$) was approximately 10% of $\lambda_{max}$ ($= \|A^T g\|_\infty$), and that setting $\lambda \approx 0.1 \times \lambda_{max}$ yielded favorable results. In this study, $\lambda$ corresponds to $\alpha_1/\mu$, and $\lambda_{max}$ is equal to the reciprocal of $\mu_0$. Hence, it is suggested from their report that the optimal $\mu/\mu_0$ value is approximately $\alpha_1 \times 10$. Although our optimal $\mu/\mu_0$ value that minimizes the mSSIM is slightly higher than that expected from their study [10], our results shown in Fig. 2 can be considered as consistent with their results.

The PRMSE is a useful index for quantitatively evaluating the convergence behavior of iterative image restoration and reconstruction methods [1, 13]. By contrast, the mSSIM is an index for measuring the similarity between two images, and it is designed to reflect the human visual system more effectively compared with measurements using the $\ell_2$ norm error, such as the PRMSE [16]. We previously reported that the mSSIM reflects the change in the visual image quality more effectively compared with the PRMSE; hence, the mSSIM is useful for evaluating the effectiveness of image restoration methods in terms of visual image quality [1]. Furthermore, we observed some discrepancies between evaluations performed using the two measures [1]. It was also observed in this study that, particularly when the noise level was 2%, the change in the mSSIM due to the difference in $\alpha_1$ was more significant than that in the PRMSE (Fig. 2), suggesting that the mSSIM is more sensitive to the difference in $\alpha_1$ than the PRMSE and is more useful for identifying the optimal parameter values.

The regularization parameter $\alpha_1$ in Eq. (2) controls the sparsity of the restored images through $\ell_1$ norm minimization, i.e., the sparsity increases with $\alpha_1$, and vice versa. As shown in Fig. 3, the blur of the restored images decreased with increasing $\alpha_1$. These results were primarily due to an increase in sparsity. Additionally, the penalty parameter, $\mu$, in Eq. (5) significantly affected the image quality of the restored images (Fig. 3). This parameter is associated significantly with the convergence behavior expressed in Eq. (5) because $\mu$ is inversely proportional to the step size in the iterative computation for solving Eq. (5). In general, when the step size in the iterative computation is extremely large, the convergence speed increases, but it does not converge to an accurate solution. Conversely, when the step size is extremely small, the convergence speed is extremely slow and cannot converge to the solution until the stopping criteria are satisfied. This appears to be the main reason for the changes in the PRMSE and mSSIM values observed when $\mu/\mu_0$ was varied (Fig. 2).

Because the $\varepsilon$ in Eq. (7) depends on the noise variance, it should be selected based on the noise level. As previously described, we set $\varepsilon$ to $\sqrt{N_{ob}}\sigma_n$ in this study. Afonso et al. [6] set this value to $\sqrt{N_{ob} + 8\sqrt{N_{ob}}}\sigma_n$. When $N_{ob}$ is large, $\sqrt{N_{ob} + 8\sqrt{N_{ob}}}\sigma_n$ can be approximated as $\sqrt{N_{ob}}\sigma_n$. In this study, $N_{ob}$ represents the total number of pixels (128 × 128). The values of $\varepsilon$ when it was expressed as $\sqrt{N_{ob}}\sigma_n$ and $\sqrt{N_{ob} + 8\sqrt{N_{ob}}}\sigma_n$ did not differ significantly. Because we assumed that the noise obeyed the Gaussian distribution in the simulation studies, it was valid to set $\sigma_n$ as a constant. However, when the noise obeys distributions other than the Gaussian distribution, such as the Poisson distribution, we cannot set $\varepsilon$ to a constant because $\sigma_n$ changes based on the image intensity [1].

## 5 Conclusion

We presented the application of the ADMM to medical image restoration and investigated its usefulness through simulation studies using a numerical phantom. Furthermore, we presented a method for generating the system matrix using a PSF and investigated the optimal parameter values using $\|A^T g\|_\infty$ as a merkmal. This study will be useful for identifying the optimal parameter values in the ADMM when applied to medical image restoration, although further studies using a wider range and greater number of medical images, as well as comparative studies with other existing methods will be necessary to fully validate the present method.

**References**

1. Murase K. New image-restoration method using a simultaneous algebraic reconstruction technique: comparison with the Richardson-Lucy algorithm. Radiol Phys Technol. 2020;13: 365–77.

2. Lustig M, Donoho D, Pauly JM. Sparse MRI: The application of compressed sensing for rapid MR imaging, Magn Reson Med. 2007;58:1182–95.

3. Daubechies I, De Friese M, De Mol C. An iterative thresholding algorithm for linear inverse problems with a sparsity constraint. Commun Pure Appl Math. 2004;57: 1413–57.

4. Beck A, Tebouble M. Fast gradient-based algorithms for constrained total variation image denoising and deblurring problems. IEEE Trans Image Process. 2009;18:2419–34.

5. Gabay D, Mercier B. A dual algorithm for the solution of nonlinear variational problems via finite element approximation. Comput Math Appl. 1976;2: 17–40.

6. Afonso MV, Bioucas-Dias JM, Figueiredo MAT. An augmented Lagrangian approach to the constrained optimization formulation of imaging inverse problems. IEEE Trans Image Process. 2011;20:681–95.

7. Tang B, Li J, Liang J. Alternating direction method of multipliers for radar waveform design in spectrally crowded environments. Signal Processing. 2018;142: 398–402.

8. Chen C, Ng MK, Zhao X-L. Alternating direction method of multipliers for nonlinear image restoration. IEEE Trans Image Process. 2015;24: 33–43.

9. Setzer S. Operator splittings, bregman methods and frame shrinkage in image processing. Int J Comput Vis. 2011:92: 265–80.

10. Boyd S, Parikh N, Chu E, Peleato B, Eckstein J. Distributed optimization and statistical learning via the alternating direction method of multipliers. Found Trends Mach Learn. 2011;3: 1–122.

11. Danielson C. An alternating direction method of multipliers algorithm for symmetric model predictive control. Optim Control Appl Meth. 2021;42: 236–60.

12. Afonso MV, Bioucas-Dias JM, Figueiredo MAT. Fast image recovery using variable splitting and constrained optimization. IEEE Trans Image Process. 2010;19:2345–56.

13. Murase K. Simultaneous correction of sensitivity and spatial resolution in projection-based magnetic particle imaging. Med Phys. 2020;47:1845–59.

14. Collins DL, Zijdenbos AP, Kollokian V, Sled JG, Kabani NJ, Holmes CJ, et al. Design and construction of a realistic digital brain phantom. IEEE Trans Med Imaging. 1998;17:463–8.
15. Murase K, Yamazaki Y, Shinohara M, Kawakami K, Kikuchi K, Miki H, et al. An anisotropic diffusion method for denoising dynamic susceptibility contrast-enhanced magnetic resonance images. Phys Med Biol. 2001;46:2713–23.
16. Wang Z, Bovik AC, Sheikh HR, Simoncelli EP. Image quality assessment: from error visibility to structural similarity. IEEE Trans Image Process. 2004;13:600–12.